\documentstyle[12pt]{article}

\begin{document}

\vspace{2mm}

\begin{flushright}
MRI-PHY/97-06 \\
hep-th/9704047
\end{flushright}

\vspace{2ex}

\begin{center}
{\large \bf 
Graviton-Dilaton Cosmology} 

\vspace{6mm}
{\large  S. Kalyana Rama}
\vspace{3mm}

Mehta Research Institute, 10 Kasturba Gandhi Marg, 

Allahabad 211 002, India. 

\vspace{1ex}
email: krama@mri.ernet.in \\ 
\end{center}

\vspace{4mm}

\begin{quote}
ABSTRACT. 
We show that the evolution of the universe is 
singularity free in a class of graviton-dilaton 
models. 

\end{quote}

\newpage

We present a general analysis of the evolution of 
a homogeneous isotropic universe in graviton-dilaton 
theories and show that the evolution of the observed 
universe is singularity free in a class of models. 
\footnote{The possibilties of such a model arising 
within string theory is discussed in \cite{k1,k3}. See 
also \cite{k5}. Other proposals for singularity free 
evolution of the universe in string theory can be found 
in \cite{vafa}.} 


\begin{center}
{\bf 1. Action and Equations of Motion} 
\end {center} 


In the absence of a dilaton potential, the most general 
graviton-dilaton action is given by 
\begin{equation}\label{schi}
S = - \frac{1}{2} \int d^4 x \sqrt{- g} \left( \chi R 
+ \frac{\omega (\chi)}{\chi} \; (\nabla \chi)^2 \right) 
+ S_M ({\cal M}, \; g_{\mu \nu})  \; , 
\end{equation}
where $\chi$ is the dilaton with the range 
$0 \le \chi \le 1$ \footnote{$\chi (today) = 1$ 
corresponding to a non zero value of the Newton's 
constant.}, $\omega (\chi)$ is the arbitrary function 
that characterises the theory. In (\ref{schi}), 
the matter fields feel only the gravitational force 
and, hence, the physical metric is $g_{\mu \nu}$ and 
the physical quantities are those directly obtained 
from $g_{\mu \nu}$. 

In the model derived in \cite{k1}, the function 
$\Omega (\chi) \equiv 2 \omega (\chi) + 3$ 
is required to satisfy 
\begin{eqnarray}
\Omega (0) & = & \Omega_0 \le \frac{1}{3}  \nonumber \\ 
\frac{d^n \Omega}{d \chi^n} & = & {\rm finite} 
\; \; \; \; \forall \; n \ge 1, \; \; \; \; 
0 \le \chi < 1  \nonumber \\ 
\Omega & \to & \infty \; \; \; \; {\rm at} 
\; \; \chi = 1 \; \; {\rm only} \nonumber \\ 
\lim_{\chi \to 1} \Omega & = & 
\Omega_1 (1 - \chi)^{- 2 \alpha} \; , \; \; \; \; 
\frac{1}{2} \le \alpha < 1 \; , \label{123} 
\end{eqnarray} 
where $\Omega_0 > 0$ and $\Omega_1 > 0$ are constants. 
We further assume, for the sake of simplicity, that 
$\Omega (\chi)$ is a strictly increasing function of 
$\chi$. The function $\Omega (\chi)$ is otherwise 
arbitrary. 

In the following, we take ``matter'' to be a perfect 
fluid with density $\rho$ and pressure $p$, related by 
$p = \gamma \rho$ where $-1 \le \gamma \le 1$ and take 
the line element to be given by 
\begin{equation}\label{ds}
d s^2 = - d t^2 + e^{2 A(t)} \left( d r^2 + r^2 
(d \theta^2 + \sin^2 \theta d \varphi^2) \right) \; . 
\end{equation} 
The equations of motion from (\ref{schi}) are 
\begin{eqnarray}
\dot{A} & = & - \frac{\dot{\chi}}{2 \chi} + \epsilon \; 
\sqrt{\frac{\rho}{6 \chi} + \frac{\Omega \dot{\chi}^2}
{12 \chi^2}} \label{ad} \\
\ddot{\chi} + 3 \dot{A} \dot{\chi} 
+ \frac{\dot{\Omega} \dot{\chi}}{2 \Omega} & = & 
\frac{(1 - 3 \gamma) \rho}{2 \Omega} \label{chidd} \\ 
\rho & = & \rho_0 e^{- 3 (1 + \gamma) A} \; , \label{rho} 
\end{eqnarray} 
where upper dots denote $t$-derivatives, 
$\epsilon = \pm 1$ and we have used $\Omega = 2 
\omega + 3$ and $p = \gamma \rho$. The square roots 
are to be taken with a positive sign always. 
Equations (\ref{ad}) and (\ref{chidd}) can also be 
written as \cite{k3} 
\begin{eqnarray}
\dot{\chi} (t) & = & \frac{e^{- 3 A}}{\sqrt{\Omega}} \; 
(\sigma (t) + c) \label{chid} \\ 
2 \chi \; \frac{d A}{d \chi} & = & - 1 
+ \epsilon \; {\rm sign} (\dot{\chi}) \; \sqrt{K} \; , 
\label{achi}
\end{eqnarray} 
where $c = \dot{\chi} \sqrt{\Omega} e^{3 A} \; 
\vert_{t = t_i}$ is a constant, $t_i$ is an initial 
time, and 
\begin{eqnarray}
\sigma (t) & \equiv & \frac{(1 - 3 \gamma) \rho_0}{2} 
\int^t_{t_i} dt \frac{e^{- 3 \gamma A}}{\sqrt{\Omega}} 
\label{sigma} \\ 
K & \equiv & \frac{\Omega}{3} \; 
\left( 1 + \frac{2 \rho_0 \chi e^{3 (1 - \gamma) A}}
{(\sigma (t) + c)^2} \right) \; .  \label{achik}
\end{eqnarray}

Note that under time reversal $t \to - t$, 
$\epsilon \to - \epsilon$ in equation (\ref{ad}), 
whereas equation (\ref{chidd}) is unchanged. Thus, 
evolution for $- \epsilon$ is same as that for 
$\epsilon$, but with the direction of time reversed. 

Note also that, in general, there will appear in 
the solutions positive non zero arbitrary constant 
factors $e^{A_0}, \; \rho_0$, and $\chi_0$ in front of 
$e^A, \; \rho$, and $\chi$ respectively. However, it 
follows from (\ref{ad}) and (\ref{chidd}) that they can 
all be set equal to $1$ with no loss of generality if 
time $t$ is measured in units of 
$\sqrt{\frac{\chi_0 e^{3 (1 + \gamma) A_0}}{\rho_0}}$. 
Therefore, in the solutions below we often assume that 
$t$ is measured in appropriate units and, hence, set 
these constants equal to $1$. Note, however, that 
the model dependent constants associated with 
$\Omega (\chi)$, such as $\Omega_0$ or $\Omega_1$ 
in (\ref{123}), cannot be set equal to $1$. 

\begin{center}
{\bf 2. Analysis of the Evolution}
\end {center} 


Our main goal is to determine whether 
the evolution of the universe in the present model 
is singular or not. The task is trivial if one can solve 
the equations of motion (\ref{ad}) - (\ref{rho}) for 
arbitrary functions $\Omega (\chi)$ (or $\psi (\phi)$). 
However, explicit solutions can be obtained only in 
special cases \cite{barrow,k1} and the methods involved 
in obtaining such solutions are inapplicable in 
the general cases of interest here. 
Also, even when applicable, the details of the solution 
tend to obscure the general features of the evolution. 
Hence, a different approach is needed which is valid 
for any matter and for any arbitrary function 
$\Omega (\chi)$ and which enables one to analyse 
the evolution for the general case and obtain its 
generic features even in the absence of explicit 
solutions. We will present such an approach below. 

Note that for the singularities to be absent, 
{\em all} curvature invariants must remain finite. 
A sufficient condition, proved in \cite{k3}, for 
{\em all} curvature invariants to be finite is 
that the quantities  
\begin{equation}\label{qty}
e^{- A}; \; \; \; 
\frac{\rho}{\chi}, \;  
\frac{\rho}{\chi \Omega}; \; \; \; 
\frac{\dot{\chi}}{\chi}, \; 
\frac{\Omega \dot{\chi}^2}{\chi^2}; \; \; \; 
{\rm and} \; \; \; 
\frac{\chi^n}{\Omega} \; \frac{d^n \Omega}{d \chi^n} \; 
\left( \frac{\dot{\chi}}{\chi} \right)^n, \; \; 
\forall n \ge 1 
\end{equation}
be all finite. To proceed, we also need initial 
values of $\dot{A}$, $\dot{\chi}$, and $\chi$ or 
equivalently $\Omega$ at initial time $t_i$. However, 
their numerical values are not needed in our approach as 
long as they are generic \footnote{A {\em non generic} 
example: $\dot{\chi} (t_i) = 0$ and $\chi (t_i) = 1$, 
{\em i.e.} $\Omega \vert_{t_i} = \infty$.} which we 
assume to be the case. 

The value $\Omega (t_i) \stackrel{<}{_\sim} 3$ 
will turn out to be special. Thus, there are 16 sets of 
possible initial conditions: 2 each for the signs of 
$\dot{A} (t_i), \; \dot{\chi} (t_i)$, and $\epsilon$, 
and 2 for whether $\Omega (t_i) \stackrel{<}{_\sim} 3$ 
or not. 

However, it is not necessary to analyse all of these 
16 sets. The evolution for $- \epsilon$ is same as that 
for $\epsilon$, but with the direction of time reversed. 
Hence, only 8 sets need to be analysed. Now, let 
$\epsilon = 1$. If $\dot{\chi} < 0$ then $\dot{A}$ can 
not be negative for any value of $\Omega$, see (\ref{ad}). 
Similarly, if $\dot{\chi} > 0$ and $\Omega > 3$ then also 
$\dot{A}$ cannot be negative since 
$\sqrt{\frac{\rho}{6 \chi} 
+ \frac{\Omega \dot{\chi}^2}{12 \chi^2}}  \ge 
\sqrt{\frac{\Omega}{3}} \; \frac{\dot{\chi}}{2 \chi} 
\; > \frac{\dot{\chi}}{2 \chi}$ in equation (\ref{ad}). 
Thus, there remain only 5 sets of possible initial 
conditions. We choose these sets so as to be of direct 
relevence to the evolution of observed universe and 
analyse each of them for $t > t_i$. We then 
use these results in section 6 to describe 
the generic evolution of the universe. We also 
assume that $\rho \ne 0$ and $\gamma \ne 1$ 
identically, as is relevent to the observed universe. 

Two remarks are now in order. First, the amount and 
the nature of the dominant ``matter'' in the universe 
are given by $\rho_0$ and $\gamma$ where $- 1 \le \gamma 
\le 1$. The later varies as the scale factor $e^A$ of 
the universe evolves. If $e^A$ is increasing, {\em i.e.} 
if $\dot{A} > 0$, then the universe is expanding, 
eventually becoming dominated by ``matter'' with 
$\gamma < \frac{1}{3}$, when such ``matter'' is present. 
For the observed universe, which is known to contain dust 
for which $\gamma = 0$, one may take $\gamma \le 0$. Thus, 
by choosing $t_i$ suitably we may assume, with no loss of 
generality, that $\gamma < \frac{1}{3}$ ($\le 0$ for 
observed universe) if $\dot{A} > 0$. Similarly, if 
$\dot{A} < 0$ then we may assume, with no loss of 
generality, that $\gamma \ge \frac{1}{3}$ when such 
``matter'' is present. This is valid for the observed 
universe which is known to contain radiation for which 
$\gamma = \frac{1}{3}$. 

Second, consider $\sigma (t)$ in (\ref{chid}). 
The integrand is positive and 
$\propto \frac{1}{\sqrt{\Omega}}$. (The dependence on 
$\rho_0$ can be  absorbed into the unit of time, as 
explained in section 3.) Hence, the value of the integral 
is controlled by $\Omega$. For example, in the limit 
$\chi \to 1$, it is controlled by the constant $\Omega_1$ 
in (\ref{123}). We will use this fact in the following. 
Also, since $t > t_i$, the sign of $\sigma (t)$ is same 
as that of $(1 - 3 \gamma)$. Moreover, $\sigma (t) = 0$ 
if $\rho_0 = 0$ or $\gamma = \frac{1}{3}$ corresponding 
to an universe containing, respectively, no matter or 
radiation only. 

\begin{center}
{\bf 2 a. $\dot{A} (t_i) > 0, \; \; 
\dot{\chi} (t_i) > 0$, and $\Omega (t_i) > 3$} 
\end {center} 


These conditions imply that $\epsilon = + 1$ in 
(\ref{ad}) and the constant $c > 0$ in (\ref{chid}). 
For $t > t_i$, the scale factor $e^A$ increases and, 
eventually, $\gamma$ can be taken to be $< \frac{1}{3}$ 
when such ``matter'' is present. In fact, $\gamma \le 0$ 
for the obeserved universe. Then, $(1 - 3 \gamma) > 0$ 
and, hence, $\sigma (t)$ increases. Thus, for $t > t_i$, 
$\dot{\chi} > 0$ and $\chi$ and $\Omega$ both increase. 
Since $\Omega > 3$, it follows from equation (\ref{ad}) 
that $\dot{A} > 0$ for $t > t_i$ and, hence, $e^A$ 
increases. Eventually, $\chi \to 1$ and 
$\Omega \to \Omega_1 (1 - \chi)^{- 2 \alpha}$. 

In the limit as $\chi \to 1, \; (\sigma (t) + c) > 0$ 
is a constant to an excellent approximation. Then, 
equation (\ref{achi}) can be solved relating $\chi$ 
and $e^A$: 
\begin{equation}\label{achi1}
e^A = {\rm (constant)} \; (1 - \chi)^{- 
\frac{2 (1 - \alpha)}{3 (1 - \gamma)}} \; . 
\end{equation} 
Substituting this result in equation (\ref{chid}) then 
yields the unique solution in the limit $\chi \to 1$: 
\begin{equation}\label{soln1}
e^A = e^{A_0} t^{\frac{2}{3 (1 + \gamma)}} 
\; , \; \; \; \; 
\chi = 1 - \chi_0 t^{- \frac{1 - \gamma}
{(1 + \gamma) (1 - \alpha)}} \; ,  
\end{equation}
where $A_0$ and $\chi_0 > 0$ are constants and $t$ is 
measured in appropriate units. Since $\chi \to 1$, it 
follows from (\ref{soln1}) that $t \to \infty$. It can 
now be seen that $(\sigma (t) + c) > 0$ is indeed 
a constant to an excellent approximation. Also, 
the quantities in (\ref{qty}) are all finite, implying 
that all the curvature invariants are finite. Thus, 
there is no singularity as $\chi \to 1$ and, thus, 
for $t > t_i$. 

\begin{center}
{\bf 2 b. $\dot{A} (t_i) > 0, \; \; 
\dot{\chi} (t_i) < 0, \; \; \Omega (t_i) > 3$, 
and $\epsilon = 1$} 
\end {center} 


The constant $c < 0$ in (\ref{chid}). For $t > t_i, \; 
\chi$ and $\Omega$ decrease and the scale factor $e^A$ 
increases. Equation (\ref{achi}) now becomes 
\[
2 \chi \; \frac{d A}{d \chi} = - 1 - \sqrt{K} \; , 
\]
where $K (t)$ is given in (\ref{achi}). 

Consider first the case $\rho = 0$ or $ \gamma = 1$, 
discussed in section 2a and which led to 
the present phase. Then, $\rho_0 (1 - 3 \gamma) \le 0$ 
and, therefore, $\sigma (t) \le 0$. Thus, 
$(\sigma (t) + c) \le c < 0$ and, hence, $\dot{\chi} (t)$ 
is negative and non zero, implying that $\chi (t)$ 
decreases for $t > t_i$. Also, $K (t)$ remains finite 
and it follows from equation (\ref{achi}) that 
$\frac{d A}{d \chi} < 0$. Thus, $\dot{A} > 0$ 
since $\dot{\chi} < 0$ and, hence, $e^A$ 
increases for $t > t_i$. 

Eventually, as $t$ increases, $\chi \to 0$ and 
$e^A \to \infty$. Also, $\frac{2 \rho_0 \chi 
e^{3 (1 - \gamma) A}}{(\sigma (t) + c)^2} \ll 1$ 
in (\ref{achi}) since $\rho_0 = 0$ or $\gamma = 1$. 
Equation (\ref{achi}) can then be solved relating 
$\chi$ and $e^A$: 
\begin{equation}\label{achi2}
e^A = {\rm (constant)} \; \chi^{- \frac{3 
+ \sqrt{3 \Omega_0}}{6}} \; , 
\end{equation}
where $\chi \to 0$ and $\Omega_0$ is given in 
(\ref{123}). Substituting this result in equation 
(\ref{chid}) then yields the unique solution in 
the limit $\chi \to 0$: 
\begin{equation}\label{soln2}
e^A = e^{A_0} (t - t_0)^n \; , \; \; \; \; 
\chi = \chi_0 (t - t_0)^m \; , 
\end{equation}
where $A_0, \; \chi_0$, and $t_0 > t_i$ are some 
constants and 
\begin{equation}\label{mn+}
n = \frac{3 + \sqrt{3 \Omega_0}}
{3 (1 + \sqrt{3 \Omega_0})} \; , \; \; \; \; 
m = \frac{- 2}{1 + \sqrt{3 \Omega_0}} \; . 
\end{equation}
Note that $m < 0$. Then, since $\chi \to 0$, it 
follows from (\ref{soln2}) that $t \to \infty$. 

Thus, as $\chi \to 0$, $t \to \infty$ and 
$e^A \to \infty$. It can also be seen that the quantities 
in (\ref{qty}) are all finite for $t_i \le t \le \infty$, 
implying that all the curvature invariants are finite. 
Thus, there is no singularity for $t_i \le t \le \infty$. 

Consider now the case where $\rho$ and $\gamma$ 
are non zero and have generic values. We have 
$\dot{A} (t_i) > 0$ and $\dot{\chi} (t_i) < 0$. Hence, 
$e^A$ increases and $\chi$ decreases. As $e^A$ increases, 
$\gamma$ can eventually be taken to be $< \frac{1}{3}$, 
when such ``matter'' is present. In fact, $\gamma \le 0$ 
for the obeserved universe. In such cases where 
$\gamma \le 0$, $\sigma (t)$ grows faster than $t$ as 
can be seen from (\ref{sigma}). Then, as $t$ increases, 
the factor $(\sigma (t) + c)$ becomes positive. 

This implies that $\dot{\chi}$, initially negative, 
passes through zero and becomes positive. We then have 
$\dot{A} > 0$ and $\dot{\chi} > 0$. As $t$ increases 
further, $\chi$ and, hence, $\Omega (\chi)$ increase. 
If $\Omega > 3$ then further evolution proceeds as 
described in section 2a. If $\Omega < 3$ then, 
upon making a time reversal, the initial conditions 
become identical to the one described in section 2e. 
The evolution in the present case then 
proceeds as in section 2d for the case which 
leads to the initial conditions of section 2e. 

\begin{center}
{\bf 2 c. $\dot{A} (t_i) < 0, \; \; \dot{\chi} (t_i) < 0$, 
and $\Omega (t_i) > 3$}
\end {center} 


These conditions imply that $\epsilon = - 1$ in 
(\ref{ad}) and the constant $c < 0$ in (\ref{chid}). 
For $t > t_i$, the scale factor $e^A$ decreases 
and, eventually, $\gamma$ can be taken to be 
$\ge \frac{1}{3}$ when such ``matter'' is present. In 
fact, this is the case for the observed universe. Then 
$(1 - 3 \gamma) \le 0$ and $\sigma (t)$ decreases 
or remains constant. Therefore, the factor 
$(\sigma (t) + c) \le c < 0$ and,  since $e^A$ decreases, 
we have that $\dot{\chi} (t) < \dot{\chi} (t_i) < 0$ for 
$t > t_i$. Thus $\chi$ and, hence, $\Omega$ decrease for 
$t > t_i$. Also, $\frac{d A}{d \chi} > 0$ since 
$\dot{\chi} < 0$ and $\dot {A} < 0$. 
 
In (\ref{achi}), $K (t_i) > 1$ since $\Omega > 3$. From 
the behaviour of $e^A, \; \chi, \; \Omega (\chi)$, and 
$(\sigma (t) + c)$ described above, it follows that $K$ 
decreases monotonically. The lowest value of $K$ is 
$\frac{\Omega_0}{3} \le \frac{1}{9}$, achievable when 
$\chi$ vanishes. This, together with $K (t_i) > 1$ 
and the monotonic behaviour of $K (t)$, then implies 
that there exists a time, say $t = t_m > t_i$ where 
$K (t_m) = 1$ with $\chi (t_m) > 0$. Hence, 
$\frac{d A}{d \chi} (t_m) = 0$. Therefore, 
$\dot{A} (t_m) = 0$ since $\dot{\chi} (t_m)$ 
is non zero. This is a critical point of $e^A$ and is 
a minimum. Also, equation (\ref{achi}) can be written as 
\begin{equation}\label{intachi}
A (t_i) - A (t_m) = \int^{\chi (t_i)}_{\chi (t_m)} 
\frac{d \chi}{2 \chi} \; (- 1 + \sqrt{K}) 
= {\rm finite} \; , 
\end{equation}
where the last equality follows because both 
the integrand and the interval of integration are 
finite. Therefore, $A (t_m)$ is finite and $> - \infty$ 
and, hence, $e^{A (t_m)}$ is finite and non vanishing. 

Thus, for $t > t_i$, the scale factor $e^A$ continues 
to decrease and reaches a non zero minimum at $t = t_m$. 
The precise values of $t_m, \; A (t_m), \; \chi (t_m)$, 
and $\Omega (t_m)$ are model dependent. Hence, nothing 
further can be said about them except, as follows from 
$K (t_m) = 1$, that $\Omega (t_m) \stackrel{<}{_\sim} 3$ 
in general and $= 3$ when $\rho_0 = 0$.  

However, the above information suffices for our purposes. 
It can now be seen that the quantities in (\ref{qty}) 
are all finite, implying that all the curvature 
invariants are finite. Thus, there is no singularity 
for $t_i \le t \le t_m$. 

For $t > t_m$, one has $\dot{A} (t) > 0, \; \dot{\chi} 
(t) < 0$, and $\Omega (\chi (t_m)) \stackrel{<}{_\sim} 
3$ by continuity. Further evolution is analysed below. 

\begin{center}
{\bf 2 d. $\dot{A} (t_i) > 0, \; \; \dot{\chi} (t_i) < 0, 
\; \; \Omega (t_i) \stackrel{<}{_\sim} 3$, and 
$\epsilon = - 1$ }
\end {center} 


This implies that the constant $c < 0$ in (\ref{chid}) 
and that $\frac{d A}{d \chi} (t_i) < 0$ in (\ref{achi}). 
Hence, $K (t_i) < 1$. For $t > t_i$, the scale factor 
$e^A$ increases and, eventually, $\gamma$ can be taken to 
be $< \frac{1}{3}$ when such ``matter'' is present. In 
fact, $\gamma \le 0$ for the observed universe. Then, 
$(1 - 3 \gamma) > 0$ and, hence, $\sigma (t)$ increases. 
Thus, $e^A$ is increases, $\chi$ decreases, and 
$(\sigma (t) + c)$ increases. Now, depending on $\rho_0$, 
$\gamma$, and the details of $\Omega (\chi)$, $K (t)$ in 
(\ref{achi}) may or may not remain $< 1$ for $t > t_i$.  

Consider first the case where $K (t)$ remains $< 1$ for 
$t > t_i$. It then follows that $(\sigma (t) + c)$ 
cannot have a zero for $t > t_i$ and, since $c < 0$, 
must remain negative and non infinitesimal. Note that 
$\gamma$ must be $> 0$ for otherwise $\sigma (t)$ grows 
as fast as or faster than $t$, making $(\sigma (t) + c)$ 
vanish at some finite time. Let 
\begin{equation}\label{ce}
\lim_{t \to \infty} (\sigma (t) + c) = c_e 
\end{equation} 
where $c_e$ is a negative, non infinitesimal constant. We 
then have for $t_i \le t \le \infty, \; c \le (\sigma (t) 
+ c) \le c_e < 0$ and, hence, $\dot{\chi} (t) < 0$. 
Therefore, eventually $\chi (t) \to 0$. Since $K (t) < 1$ 
and $\dot{\chi} (t) < 0$, it also follows from 
(\ref{achi}) that $\frac{d A}{d \chi} < 0$ and, hence, 
$\dot{A} (t) > 0$. We will now consider the limit 
$\chi \to 0$. 

$K (t) < 1$ implies that as $\chi \to 0$, 
$K \to \frac{\Omega_e}{3} < 1$ where 
\begin{equation}\label{omegae}
\Omega_e \equiv \Omega_0 \; \left( 1 
+ \frac{2 \rho_0}{c_e^2} \; \lim_{\chi \to 0} 
\chi e^{3 (1 - \gamma) A} \right) 
\end{equation} 
is a constant and $\Omega_0$ is given in (\ref{123}). 
Now, as $\chi \to 0$, equation (\ref{achi}) 
can be solved relating $\chi$ and $e^A$: 
\begin{equation}\label{achi0}
e^A = ({\rm constant}) \; 
\chi^{- \frac{3 - \sqrt{3 \Omega_e}}{6}} \; . 
\end{equation}
Note that $(3 - \sqrt{3 \Omega_e}) > 0$ since 
$\Omega_e < 3$. Hence, $e^A \to \infty$ as 
$\chi \to 0$. Substituting (\ref{achi0}) in equation 
(\ref{chid}) then yields the unique solution in 
the limit $\chi \to 0$: For $\Omega_e \ne \frac{1}{3}$, 
\begin{equation}\label{chi0}
e^A = e^{A_0} \left( t_0 - {\rm sign} (m) t \right)^n 
\; , \; \; \; \; \chi = \chi_0 
\left( t_0 - {\rm sign} (m) t \right)^m \; , 
\end{equation}
where $t$ is measured in appropriate units, $A_0, \; 
\chi_0$, and $t_0 > t_i$ are some constants, and 
\begin{equation}\label{mn} 
n = \frac{3 - \sqrt{3 \Omega_e}}
{3 (1 - \sqrt{3 \Omega_e})} \; , \; \; \; \; 
m = \frac{- 2}{1 - \sqrt{3 \Omega_e}} \; . 
\end{equation}
For $\Omega_e = \frac{1}{3}$, 
\begin{equation}\label{chi0e}
e^A = e^{A_0} e^{- \frac{c_e (t - t_0)}{3}} \; , 
\; \; \; \; \chi = \chi_0 e^{c_e (t - t_0)} \; , 
\end{equation}
where $c_e$ is defined in (\ref{ce}). 

Let $\Omega_e > \frac{1}{3}$. Hence, $m > 0$. Also, 
$n < 0$ because $\Omega_e < 3$. Since $\chi \to 0$, it 
follows from (\ref{chi0}) that $t \to t_0 > t_i$, which 
implies that $\chi$ vanishes at a finite time $t_0$. In 
this limit, the scale factor $e^A \to \infty$ since 
$n < 0$. The quantities in (\ref{qty}), for example 
$\frac{\dot{\chi}}{\chi}$, also diverge, implying that 
the curvature invariants, including the Ricci scalar, 
diverge. Thus, for $\Omega_e > \frac{1}{3}$, there is 
a singularity at a finite time $t_0$. 

Let $\Omega_e \le \frac{1}{3}$, which is the case in our 
model, see (\ref{123}). When $\Omega_e < \frac{1}{3}$, 
$m < 0$. Also, $n > 0$ because $\Omega_e < 3$. Since 
$\chi \to 0$, it follows from (\ref{chi0}) that 
$t \to \infty$. In this limit, the scale factor 
$e^A \to \infty$ since $n > 0$ now. The same result 
holds also for the solution in (\ref{chi0e}) with 
$\Omega_e = \frac{1}{3}$. 

For $K (t)$ to be $< 1$, $(\sigma (t) + c)$ must not be 
too small. Since $c < 0$ and $\sigma (t) > 0$, it 
necessarily implies that $\sigma (t)$ must remain finite. 
It follows from the above solutions that $\sigma (t)$ can 
remain finite only if $\sqrt{3 \Omega_e} > 
\frac{1 - 3 \gamma}{1 - \gamma}$, equivalently, $\gamma 
> \frac{1 - \sqrt{3 \Omega_e}}{3 - \sqrt{3 \Omega_e}}$. 
Under this condition, it can be checked easily that 
$\lim_{\chi \to 0} \chi e^{3 (1 - \gamma) A} = 0$.  
Hence, $\Omega_e = \Omega_0$ as follows from 
(\ref{omegae}). 

It can now be seen, for $\Omega_e = \Omega_0 \le 
\frac{1}{3}$, that the quantities in (\ref{qty}) are 
all finite for $t_i \le t \le \infty$, implying that 
all the curvature invariants are finite. Thus, there 
is no singularity for $t_i \le t \le \infty$ when 
$\Omega_0 \le \frac{1}{3}$.  

Consider the second case where $K (t)$ in (\ref{achi}) 
may not remain $< 1$ for all $t > t_i$. This is the case 
for the observed universe where $\gamma \le 0$ and, 
hence, $\sigma (t)$ grows as fast as or faster than $t$. 
Then there is a time, say $t = t_1$, when the value of 
$K = 1$. It follows that $c < (\sigma(t_1) + c) < 0$ and, 
hence, $\dot{\chi} (t) < 0$ for $t_i \le t \le t_1$. 
Also, $\chi (t_1)$ is non vanishing since the case of 
$\chi (t_1) \to 0$ is same as the case described above. 
Therefore, we have that $\frac{d A}{d \chi} (t_1) = 0$ 
which implies, since $\dot{\chi} (t_1) \ne 0$, that 
$\dot{A} (t_1) = 0$. This is a critical point of $e^A$ 
and is a maximum. 

For $t > t_1, K (t)$ becomes $> 1$ and, hence, 
$\dot{A} (t) < 0$. Also, $\dot{\chi} < 0$ and 
$\Omega (t) < 3$. Then $\epsilon = -1$ necessarily. 
Further evolution is analysed below. 

\begin{center}
{\bf 2 e. $\dot{A} (t_i) < 0, \; \; 
\dot{\chi} (t_i) < 0$, and $\Omega (t_i) < 3$} 
\end {center} 


The above conditions imply that $\epsilon = - 1$, 
the constant $c < 0$ in (\ref{chid}), and  
$\frac{d A}{d \chi} (t_i) > 0$. Hence, $K (t_i) > 1$. 
For $t > t_i$, $(\sigma (t) + c)$ which is negative 
at $t_i$ may or may not vanish for non zero $\chi$. 

Consider the case where $(\sigma (t) + c)$ does not 
vanish and remains negative for non zero $\chi$. 
Therefore, $\dot{\chi} (t) < 0$ and, hence, $\chi (t)$ 
decreases. For $t > t_i$, the scale factor $e^A$ 
decreases and, eventually, $\gamma$ can be taken to be 
$\ge \frac{1}{3}$ when such ``matter'' is present. In 
fact, this is the case for the observed universe. 
Then, $(1 - 3 \gamma) < 0$ and $\sigma (t)$ decreases. 
Hence, $(\sigma (t) + c)$ also decreases. 

It follows from (\ref{achi}) that $K$, initially $> 1$, 
is now decreasing since $\chi$ and $e^A$ are decreasing 
and $(\sigma (t) + c)^2$ is increasing. Its lowest value 
$= \frac{\Omega_0}{3} \le \frac{1}{9}$, achieveable at 
$\chi = 0$. Therefore, there exists a time, say 
$t = t_M$, where $K = 1$ and, hence, 
$\frac{d A}{d \chi} = 0$. Clearly, $\chi (t_M)$ and 
$\dot{\chi} (t_M)$ are non zero for the same reasons as 
in section 2c, implying that $\dot{A} (t_M) = 0$. 
This is a critical point of $e^A$ and is a minimum. 

The existence of this critical point of $e^A$ can be seen 
in another way also. As $(\sigma (t) + c) < 0$ grows in 
magnitude, it follows from equation (\ref{chid}) that 
$\Omega \dot{\chi}^2 \propto e^{- 6 A}$, whereas $\rho$ 
is given by (\ref{rho}). Note that $\dot{\chi} (t) < 0$ 
and, hence, $\chi (t)$ decreases and eventually $\to 0$. 
Then, in the limit $\chi \to 0$, 
the $\frac{\Omega \dot{\chi}^2}{\chi^2}$ term in equation 
(\ref{ad}) dominates $\frac{\rho}{\chi}$ term. Hence, in 
this limit where 
$\Omega \to \Omega_0 \le \frac{1}{3} < 3$, we have that 
$\dot{A} > 0$. Since $\dot{A} < 0$ initially, this 
implies the existence of a zero of $\dot{A}$. This is 
a critical point of $e^A$ and is a minimum. 

Note that all quantities remain finite for 
$t_i \le t \le t_M$. In particular, the quantities 
in (\ref{qty}) are all finite, implying that all 
the curvature invariants are finite. Thus, there 
is no singularity for $t_i \le t \le t_M$. 

For $t > t_M$, we have $\dot{A} > 0, \; \dot{\chi} 
< 0, \; \Omega < 3$, and $\epsilon = -1$. Further 
evolution then proceeds as described in section 2d. 
The evolution can thus become repetetive and oscillatory. 

Consider now the case where $(\sigma (t) + c)$ vanishes,
say at $t = t_{m'}$, for non zero $\chi$. Hence, 
$\dot{\chi} (t_{m'})$ vanishes and this is a minimum 
of $\chi$. For $t > t_{m'}, \; (\sigma (t) + c)$ and 
$\dot{\chi}$ are positive by continuity and, hence, 
$\chi$ increases. Also, $\dot{A} < 0$ and the scale 
factor $e^A$ decreases. Eventually, $\gamma$ can be 
taken to be $\ge \frac{1}{3}$ when such ``matter'' is 
present. In fact, this is the case for the observed 
universe. Then, $(1 - 3 \gamma) < 0$ and $\sigma (t)$ 
decreases. Hence, $(\sigma (t) + c)$ also decreases. 

Now, as $\chi$ increases and $\to 1$, $(\sigma (t) + c)$ 
may or may not vanish with $\chi < 1$.  If 
$(\sigma (t) + c)$ vanishes with $\chi < 1$, then 
$\dot{\chi}$ vanishes. This is a critical point of $\chi$ 
and is a maximum, beyond which we have $\dot{A} < 0, \; 
\dot{\chi} < 0$, and $\epsilon = -1$. Further evolution 
then proceeds as described in section 2c 
if $\Omega > 3$, and as in section 2e if 
$\Omega < 3$. The evolution can thus become repetetive 
and oscillatory.  

If $(\sigma (t) + c)$ does not vanish and remain 
positive for $\chi \le 1$ then, eventually,
$\chi \to 1$. It follows from equation (\ref{chid}) 
that $\Omega \dot{\chi}^2 \propto e^{- 6 A}$, 
whereas $\rho$ is given by (\ref{rho}). Note that 
$\dot{A} < 0$ and $e^A$ is decreasing. Then, in this 
limit, the $\frac{\Omega \dot{\chi}^2}{\chi^2}$ term in 
equation (\ref{ad}) dominates or, if $\gamma = 1$, is 
of the same order as the $\frac{\rho}{\chi}$ term. 

Thus we have $\dot{A} < 0, \; \dot{\chi} > 0, \; \chi 
\to 1$, equivalently $\Omega \to \infty$, and $\epsilon 
= - 1$. Also, $\frac{\Omega \dot{\chi}^2}{\chi^2}$ term 
in equation (\ref{ad}) dominates or, if $\gamma = 1$, is 
of the same order as the $\frac{\rho}{\chi}$ term. 
But this is precisely the time reversed version 
of the evolution analysed in section 2b and 
in section 2a for the case which led to 
the initial conditions of section 2a, where 
the dynamics is clear in terms of $\phi$ and 
$\psi (\phi)$. Applying the results of sections 2a 
and 2b, it follows that $\phi$ will cross 
the value $0$ after which $e^{\psi}$ and, hence, $\chi$ 
begins to decrease. We then have $\dot{A} < 0$, 
$\dot{\chi} < 0$, and $\Omega > 3$. Also, 
$\epsilon = - 1$ necessarily. Further evolution then 
proceeds as described in section 2c. The evolution 
can thus become repetetive and oscillatory. 

\begin{center}
{\bf 3. Evolution of observed Universe} 
\end {center} 


We now use the results of the analysis in section 2 
to describe the generic evolution of the observed 
universe. Note that the observed universe certainly 
contains dust ($\gamma = 0$) and radiation 
($\gamma = \frac{1}{3}$). As follows from (\ref{rho}), 
``matter'' with larger $\gamma$ dominates the evolution 
for smaller value of $e^A$ and vice versa. Therefore, 
when $e^A$ is decreasing we take $\gamma \ge \frac{1}{3}$ 
eventually, and when $e^A$ is increasing we take 
$\gamma \le 0$ eventually. We 
start with an initial time $t_i$, corresponding to 
a temperature, say $\stackrel{>}{_\sim} 10^{16}$ GeV,  
such that GUT symmetry breaking, inflation, and other 
(matter) model dependent phenomena may occur for 
$t > t_i$ only. Our observed universe is expanding at 
$t_i$ and, hence, $\dot{A} (t_i) > 0$. For the sake of 
definiteness, we take $\dot{\chi} (t_i) > 0$ and 
$\Omega (t_i) > 3$, equivalently $\omega (t_i) > 0$, 
as commonly assumed. 

We first describe the evolution for $t > t_i$ taking, as 
initial conditions $\dot{A} (t_i) > 0, \; \dot{\chi} 
(t_i) > 0$, and $\Omega (t_i) > 3$. Then, $\epsilon = 1$ 
necessarily. To describe the evolution for $t < t_i$, 
we reverse the direction of time and take, as initial 
conditions, $\dot{A} (t_i) < 0, \; \dot{\chi} (t_i) < 0$, 
and $\Omega (t_i) > 3$. Then, $\epsilon = - 1$ 
necessarily. The required evolution is that for $t > t_i$ 
in terms of the reversed time variable, which is also 
denoted as $t$. The results of section 2 can then 
be applied directly. 

\begin{center}
{\bf $t > t_i$} 
\end{center} 


Initially, we have $\dot{A} (t_i) > 0, \; \dot{\chi} 
(t_i) > 0$, and $\Omega (t_i) > 3$. Then, $\epsilon = 1$ 
necessarily. For $t > t_i$, the evolution proceeds as 
described in section 2a. Both $e^A$ and $\chi$ 
increase. Hence, $\Omega$ increases. Eventually, as 
$e^A$ increases, we can take $\gamma \le 0$. Then, 
as $t \to \infty$, $e^A$ and $\chi$ evolve as given 
in (\ref{soln1}). 

In particular, it can be seen that the quantities in 
(\ref{qty}) are all finite for $t_i \le t \le \infty$, 
implying that all the curvature invariants are finite. 
Hence, the evolution is singularity free. 

Thus, in the present day universe in our model, 
$\chi (today) \to 1$ and $\Omega (today) \to \infty$. 
Also, as follows from (\ref{123}), 
$\frac{1}{\Omega^3} \frac{d \Omega}{d \chi} (today) 
= - \frac{2 \alpha}{\Omega_1^2} 
(1 - \chi)^{4 \alpha - 1} \to 0$. Therefore, 
our model satisfies the observational constraints 
imposed by solar system experiments, namely 
$\Omega (today) > 2000$ and $\frac{1}{\Omega^3} 
\frac{d \Omega}{d \chi} (today) < 0.0002$. 

\begin{center}
{\bf $t < t_i$, equivalently $(- t) > (- t_i)$} 
\end{center} 


To describe the evolution for $t < t_i$, we reverse 
the direction of time. Then, in terms of the reversed 
time variable, also denoted as $t$, we have 
$\dot{A} (t_i) < 0$, $\dot{\chi} (t_i) < 0$, and 
$\Omega (t_i) > 3$. Then, $\epsilon = - 1$ necessarily. 
For $t > t_i$, the evolution proceeds as described in 
section 2c. Both $e^A$ and $\chi$ decrease. Hence, 
$\Omega$ decreases. Then, as shown in section 2c, 
there exists a time, say $t = t_m > t_i$ where 
$K (t_m) = 1$ in equation (\ref{achi}), 
$e^{A (t_m)} > 0$, $\chi (t_m) > 0$, and 
$\dot{\chi} (t_m) < 0$. Note that $\Omega (t_m) < 3$. 

As shown in section 2c, $K (t_m) = 1$ 
implies that $\dot{A} (t_m) = 0$. Hence, the scale 
factor $e^A$ reaches a minimum. For $t > t_m$, we then 
have $\dot{A} > 0$, $\dot{\chi} < 0$, $\Omega < 3$, and 
$\epsilon = - 1$. The evolution proceeds as described in 
section 2d. Now, however, the evolution for 
$t > t_m$ is complicated since the universe is known to 
contain ``matter'' with $\gamma = 0$. Nevertheless, all 
qualitative features of its evolution can be obtained 
using the results of section 2. 

For $t \stackrel{>}{_\sim} t_m$, $\dot{A} > 0$, 
$\dot{\chi} < 0$, and $(\sigma (t) + c) < 0$. Hence, 
$e^A$ increases and $\chi$ decreases and, eventually, 
``matter'' with $\gamma \le 0$ dominates the evolution. 
Then $(1 - 3 \gamma) > 0$ and $\sigma (t)$ begins to 
increase. Hence $(\sigma (t) + c)$, which is negative, 
also begins to increase. For $\gamma \le 0$, $\sigma (t)$ 
grows atleast as fast as $t$, as follows from 
(\ref{sigma}). Then eventually, as described in section 
2d, $e^A$ reaches a maximum eventually at time, 
say $t = t_1$. Also, $(\sigma (t_1) + c) < 0$ and, hence, 
$\dot{\chi} (t_1) < 0$. Note that this critical point of 
$e^A$ exists independent of the details of the model, as 
long as ``matter'' with $\gamma \le 0$ is present. Further 
evolution then proceeds as described in section 2e. 

For $t \stackrel{>}{_\sim} t_1$, we then have $\dot{A} 
< 0, \; \dot{\chi} < 0$, and $\Omega < 3$. Also, 
$(\sigma (t) + c) < 0$ but $\sigma (t)$ remains 
increasing. Hence, depending now on the details of 
the model, $(\sigma (t) + c)$ may remain negative for all 
$t > t_1$, or it may reach a zero and become positive. 

In the first case, $(\sigma (t) + c)$ remains negative 
for all $t > t_1$. Hence, $\dot{\chi} < 0$ and $\chi$ 
continues to decrease. The scale factor $e^A$, as shown 
in section 2e, reaches a minimum at time, say 
$t = t_M > t_1$, with $\chi (t_M) > 0$ non vanishing. 
For $t > t_M, \; e^A$ increases, and we have $\dot{A} > 0, 
\; \dot{\chi} < 0$, and $\Omega < 3$. Further evolution 
then proceeds as described in section 2d. 

Note that this means that the scale factor increases and 
reaches a maximum, then decreases and reaches a minimum, 
then increases and so on. The existence of maxima of 
$e^A$ is model independent as long as ``matter'' with 
$\gamma \le 0$ is present, which is the case for 
the observed universe. The minima of $e^A$ are all non 
zero for the reasons described in section 2c. 
Their existence depends on whether $(\sigma (t) + c)$ 
remains negative or not, but is otherwise model 
independent as long as ``matter'' with 
$\gamma \ge \frac{1}{3}$ is present, which is the case 
for the observed universe. The evolution can thus become 
repetetive and oscillatory. 

In the second case, $(\sigma (t) + c)$ reaches a zero 
at time, say $t = t_{m'} > t_1$ and becomes positive. 
It then follows that $\dot{\chi} (t_{m'}) = 0$ and that 
this is a minimum of $\chi$. For $t > t_{m'}$, we have 
$\dot{A} < 0$, $\dot{\chi} > 0$, and $\Omega < 3$. 
Hence, $e^A$ decreases and $\chi$ increases. 
Eventually, we can take $\gamma \ge \frac{1}{3}$. 

Now, $(1 - 3 \gamma) \le 0$ and $\sigma (t)$ begins to 
decrease or remains constant. Hence $(\sigma (t) + c)$, 
which is positive, also begins to decrease or remains 
constant. Thus, depending on the details of $\rho_0$ 
and $\gamma$, $(\sigma (t) + c)$ may or may not vanish 
with $\chi < 1$. If $(\sigma (t) + c)$ vanishes with 
$\chi < 1$, then $\dot{\chi}$ vanishes. This critical 
point is a maximum of $\chi$, beyond which we have 
$\dot{A} < 0$, $\dot{\chi} < 0$, and $\epsilon = -1$. 
Further evolution then proceeds as described in section 2c 
if $\Omega > 3$ and as in section 2e if 
$\Omega < 3$. The evolution can thus become repetetive 
and oscillatory. 

If $(\sigma (t) + c)$ does not vanish and remain 
positive for $\chi < 1$ then, eventually, $\chi \to 1$. 
It follows from equation (\ref{chid}) that 
$\Omega \dot{\chi}^2 \propto e^{- 6 A}$, whereas $\rho$ 
is given by (\ref{rho}). Note that $\dot{A} < 0$ and 
$e^A$ is decreasing. Then, in this limit, 
the $\frac{\Omega \dot{\chi}^2}{\chi^2}$ term in 
equation (\ref{ad}) dominates or, if $\gamma = 1$, 
is of the order of the $\frac{\rho}{\chi}$ term. 

Thus we have $\dot{A} < 0$, $\dot{\chi} > 0$, 
$\chi \to 1$, equivalently $\Omega \to \infty$, 
and $\epsilon = - 1$. Also, 
$\frac{\Omega \dot{\chi}^2}{\chi^2}$ term in equation 
(\ref{ad}) dominates or is of the order of 
the $\frac{\rho}{\chi}$ term. But this is precisely 
the time reversed version of the evolution analysed in 
section 2b, where the dynamics is clear in terms 
of $\phi$ and $\psi (\phi)$. Applying the results of 
section 2b, it then follows that $\phi$ will cross 
the value $0$ after which $e^{\psi}$ and, hence, $\chi$ 
will begin to decrease. We then have $\dot{A} < 0$, 
$\dot{\chi} < 0, \Omega > 3$. The evolution then 
proceeds as described in section 2c. The 
evolution can thus become repetetive and oscillatory. 

Thus it is clear that depending on the details of 
the model, the universe undergoes oscillations, perhaps 
infinitely many. As follows from the above description, 
the oscillations can stop, if at all, only in the limit 
$\chi \to 0$. The solutions then are given by 
(\ref{chi0}) - (\ref{chi0e}). In particular, however, 
the quantities in (\ref{qty}) all remain finite during 
the oscillations, implying that all the curvature 
invariants remain finite. They also remain finite 
in the solutions (\ref{chi0}) - (\ref{chi0e}) 
if $\Omega_0 \le \frac{1}{3}$ in (\ref{123}). 

Thus, it follows that if $\Omega_0 \le \frac{1}{3}$ 
then the quantities in (\ref{qty}) are all finite for 
$t_i \le t \le \infty$, implying that all the curvature 
invariants are finite. Hence, the evolution is 
singularity free. 

In summary, the evolution of a realistic universe, such 
as our observed one, with the initial conditions 
$\dot{A} (t_i) > 0$, $\dot{\chi} (t_i) > 0$, and 
$\Omega (t_i) > 3$ proceeds in the present model as 
follows. For $t > t_i$, the scale factor $e^A$ increases 
continuously to $\infty$. The field $\chi$ increases 
continuously to $1$. Correspondingly, $\Omega$ increases 
continuously to $\infty$. 

For $t < t_i$, the scale factor $e^A$ decreases and 
reaches a non zero minimum. It then increases and 
reaches a maximum, then decreases and reaches a minimum, 
then increases and so on, perhaps ad infintum. 
The oscillations can stop, if at all, only in the limit 
$\chi \to 0$. The solutions then are given by 
(\ref{chi0}) - (\ref{chi0e}). The field $\chi$ 
may, depending on the details of model, continuously 
decrease to $0$, or undergo oscillations, its maxima 
always being $\le 1$. 

Also, the curvature invariants all remain finite for 
$- \infty \le t \le \infty$. Hence, the evolution, 
although complicated and model dependent in details, is 
completely singularity free if $\Omega (\chi)$ satisfies 
(\ref{123}). Thus we have that a homogeneous isotropic 
universe, such as our observed one, evolves with no big 
bang or any other singularity in a class of 
models where $\Omega (\chi)$ satisfies (\ref{123}). 
The time continues indefinitely into the past and 
the future, without encountering any singularity. 

When the initial conditions are different, the evolution 
can again be analysed along similar lines. However, 
the main result that the evolution is singularity free 
remains unchanged. 


\end{document}